\documentclass{webofc}
\usepackage[varg]{txfonts}   
\usepackage{hyperref}
\usepackage[mathlines]{lineno}
\begin{document}
\title{Machine Learning for Columnar High Energy Physics Analysis}
%
%

\author{\firstname{Elliott} \lastname{Kauffman}\inst{1}\fnsep\thanks{\email{ek8842@princeton.edu}} \and
        \firstname{Alexander} \lastname{Held}\inst{2}\fnsep \and
        \firstname{Oksana} \lastname{Shadura}\inst{3}\fnsep
}

\institute{Princeton University 
\and
           University of Wisconsin-Madison 
\and
           University of Nebraska-Lincoln
          }

\abstract{%
  Machine learning (ML) has become an integral component of high energy physics data analyses and is likely to continue to grow in prevalence. 
  Physicists are incorporating ML into many aspects of analysis, from using boosted decision trees to classify particle jets to using unsupervised learning to search for physics beyond the Standard Model. 
  Since ML methods have become so widespread in analysis and these analyses need to be scaled up for HL-LHC data, neatly integrating ML training and inference into scalable analysis workflows will improve the user experience of analysis in the HL-LHC era.\\ \\
  We present the integration of ML training and inference into the IRIS-HEP Analysis Grand Challenge (AGC) pipeline to provide an example of how this integration can look like in a realistic analysis environment. 
  We also utilize Open Data to ensure the project’s reach to the broader community. 
  Different approaches for performing ML inference at analysis facilities are investigated and compared, including performing inference through external servers. 
  Since ML techniques are applied for many different types of tasks in physics analyses, we showcase options for ML integration that can be applied to various inference needs.
}
\maketitle
\section{Introduction}
\label{intro}
Machine learning has increased in prevalence across many areas in high energy physics, a trend which is likely to continue through the high luminosity era of the LHC (HL-LHC). 
The large luminosity increase initiated by this upgrade will introduce much higher data volume and event size. 
Machine learning is well-positioned to help with many of the resulting challenges ~\cite{albertsson_kim_2019}. 
Analysis in high energy physics is one area where machine learning promises to assist, including tasks such as event classification, jet tagging, and unsupervised anomaly detection.\\ \\
Traditionally high energy physicists have depended on compiled languages such as C++ due to the need for fast processing speed and efficient memory usage. 
Analyses have been typically structured in "event-loop" format, in which event-level quantities are calculated and handled in a for-loop over events. 
An alternative to this approach is columnar analysis, in which events are processed in array format where events are rows and their associated quantities are columns. 
This allows one to avoid writing explicit for-loops and exploit parallel computations. 
Columnar analysis pipelines also have the benefit of higher readability and provide the ability to develop an analysis interactively. 
These features combine to offer a faster time-to-insight for physicists performing analysis ~\cite{pivarski_jim_2020} ~\cite{smith_nick_2020}.\\ \\
Columnar analyses are also popular in other fields, allowing HEP columnar analyses to utilize well-maintained tools that are broadly used outside of HEP. 
This may be especially useful when it comes to machine learning, since Python offers an extensive and well-maintained collection of machine learning tools. 
We present a columnar high energy physics analysis pipeline with integrated machine learning inference in the context of the IRIS-HEP Analysis Grand Challenge.
\section{The Analysis Grand Challenge}
\label{agc-background}
The Institute for Research and Innovation in Software for High Energy Physics (IRIS-HEP)~\cite{iris-hep} develops software and software infrastructure in anticipation of the HL-LHC. 
The Analysis Grand Challenge~\cite{held_alexander_2022} integrates the different components of the IRIS-HEP ecosystem into realistic HEP analysis workflows involving HL-LHC-scale data. 
The project can serve multiple purposes, including testing the integration of various tools, serving as a test for different tools and analysis facilities, and introducing users to columnar HEP analysis. 
At the time of this paper, the most developed analysis workflow under the AGC umbrella is a $t\bar{t}$ cross-section measurement using CMS 2015 Open Data~\cite{cms-open-data}. 
This particular task was chosen so that all typical analysis workflow aspects can be reflected.
The files used for this task, consisting of a total of 1 billion events (around 1.8 TB), are pre-converted into NanoAOD format.\\ \\
The IRIS-HEP implementation of the Analysis Grand Challenge and information for using the relevant input data is located on GitHub~\cite{agc_code}. 
More information about the Analysis Grand Challenge can be found on the dedicated Read the Docs~\cite{agc_rtd}.
\subsection{CMS Open Data Task Description}
\label{task-description}
The IRIS-HEP implementation of the task can be ran fully within \texttt{Jupyter} notebooks, allowing the user to develop their analysis in an interactive manner. 
Before the introduction of the machine learning component, the $t\bar{t}$ cross-section measurement task can be broken down into the following steps. 
First, the appropriate data are accessed and events are selected based on cuts expected to improve the signal region. 
At the same time, some object and event-level variations are applied for the purpose of systematic uncertainties. The tools used to handle this step in this implementation are \texttt{uproot}~\cite{uproot}, \texttt{coffea}~\cite{coffea}, and \texttt{awkward}~\cite{awkward}. 
There is an optional step to pre-select relevant columns and save these in cached files using \texttt{ServiceX}~\cite{servicex}. 
Next, the relevant observables are calculated and filled into histograms using \texttt{hist}~\cite{hist}. 
These histograms are saved to \texttt{ROOT}~\cite{root} files. 
The histograms are then used to build statistical models and create a workspace. 
A fit is performed using pseudodata generated from simulated data to measure the value of the $t\bar{t}$ cross-section. The tools used for this step are \texttt{pyhf}~\cite{pyhf}~\cite{pyhf_joss} and \texttt{cabinetry}~\cite{cabinetry}.\\ \\
\subsection{Machine Learning Task}
\label{ml-task}
\begin{figure}[h]
\centering
\includegraphics[width=10cm,clip]{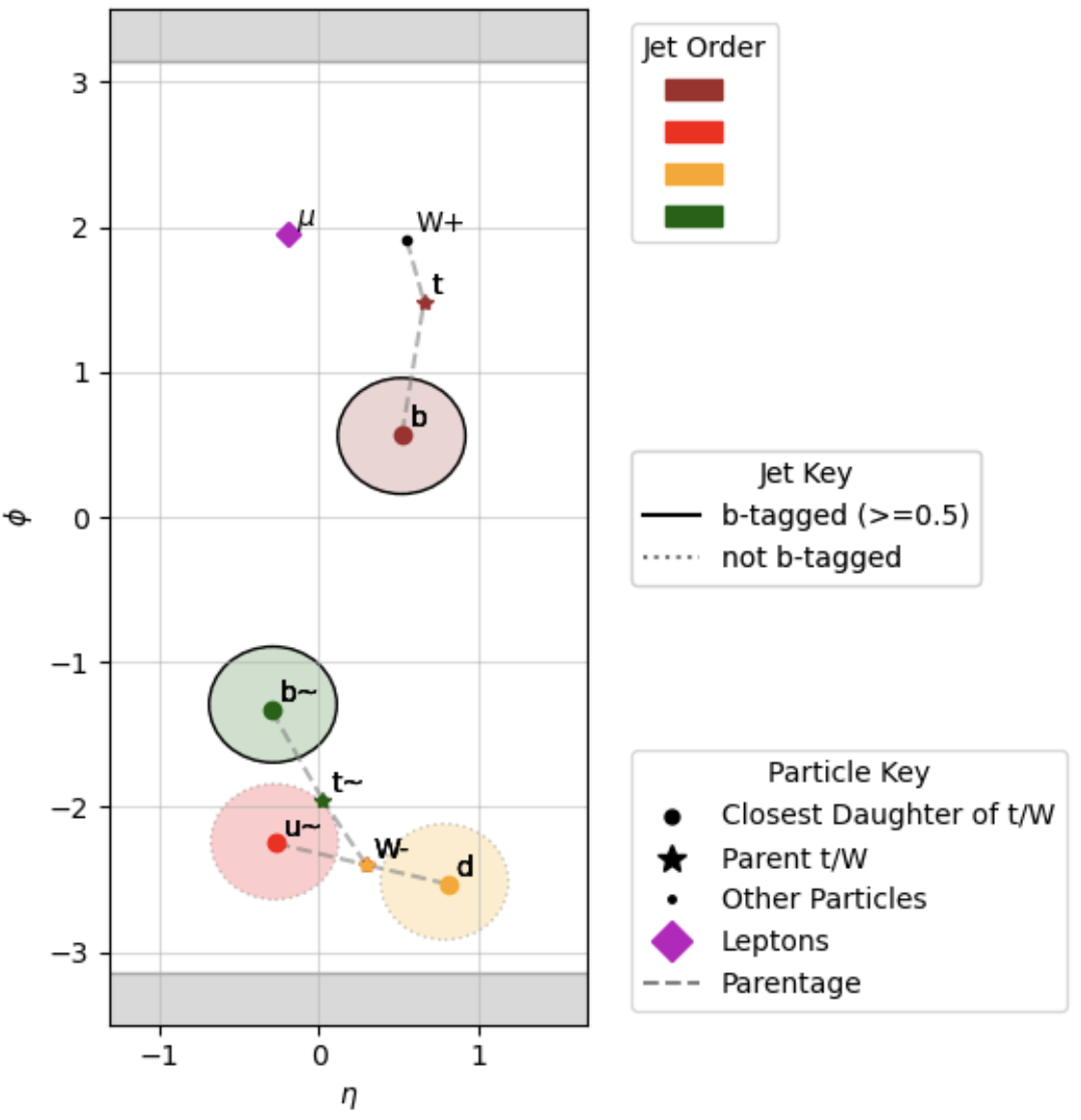}
\caption{An example of a four-jet $t\bar{t}$ event in $\eta$-$\phi$ space with two $b$-tagged jets outlined in black. Truth particles are also displayed with parentage indicated by dashed grey lines. Jets are ordered by decreasing transverse momentum.}
\label{ttbar_example}
\end{figure}

The goal of the machine learning task implemented in this first iteration is to use a simple model as a proof of principle. 
One of the observables calculated in the existing $t\bar{t}$ pipeline is a reconstructed mass of the top quark. 
The signal used for this search contains four jets and one lepton: one top quark decays to a jet and a charged lepton plus a neutrino while the other decays to three jets. \\ \\
\begin{figure}[h]
\centering
\includegraphics[width=6cm,clip]{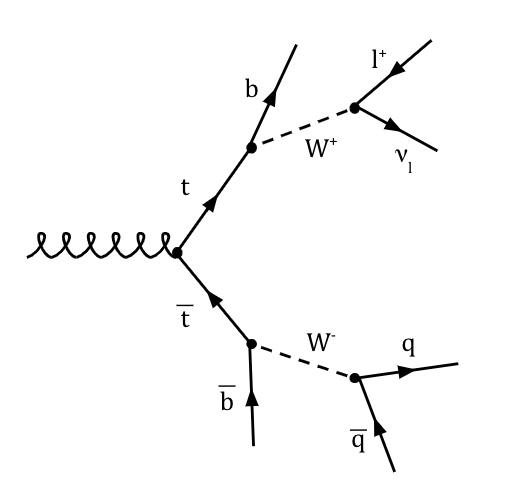}
\caption{Feynman diagram representing the $t\bar{t}$ signal used for the cross-section measurement (4 jets and a charged lepton plus a neutrino)}
\label{ttbar_diagram}       
\end{figure}
The three jets with the highest combined transverse momentum ($p_T$) correlate highly with the three jets on the side of hadronic decay, so the combined mass of these three jets should approximate the mass of the top quark. 
With an approach based on a boosted decision tree (BDT), this reconstruction can be more accurate. 
In addition to the increased accuracy of this observable, this jet-labelling unlocks access to other observables. 
For instance, if one knows with good certainty which two jets originate from the $W$ boson, one can estimate the $W$ mass.
One could also estimate the angles between the different jets using their predicted labels.\\ \\
The labeling scheme is as follows: the two jets originating from the decay products of the W boson are considered indistinguishable and labeled as $W$. 
The jet originating from the $b$ quark on the side of leptonic decay is called $b_{\text{top-had}}$. 
The jet originating from the $b$ quark on the side of hadronic decay is called $b_{\text{top-lep}}$. 
The BDT considers each permutation of jet assignments within an event as an input. 
The permutation with the highest BDT score is selected and the jets are labeled according to that permutation.
One example of a simulated event containing the required labels is shown in Fig. \ref{ttbar_example}.\\ \\
\pagebreak
\section{Training Component}
The machine learning task is implemented within the $t\bar{t}$ cross-section measurement task using a pre-trained model so that only machine learning inference is required. 
A separate pipeline is used to produce the trained model. 
First, simulated data is loaded from \texttt{ROOT} files using \texttt{uproot}. 
Each event is filtered to ensure that it fits the signal region (requiring 4 jets and 2 $b$-tagged jets). 
The events are further filtered using the truth information of particles which are matched to jets using a standard $\Delta R$ method. 
Events that contain exactly two $W$-labeled jets, one $b_{\text{top-had}}$-labeled jet, and one $b_{\text{top-lep}}$-labeled jet are selected from this sample. 
For the training dataset it is necessary that perfect jet-parton assignment is possible. 
Since not many events are required to produce a reasonable trained model, restricting to four jets per event is convenient to improve computational efficiency. 
For the four jet case, there are 12 different permutations of jet-label associations to consider. \\ \\
For each possible permutation, 20 features are calculated. 
These features include observables such as the $\Delta R$ between the $b_{\text{top-lep}}$ jet and the lepton and the transverse momentum of each jet. If $N$ is the number of events used for training, the input training dataset has dimensions $(12\times N, 20)$. 
These columns of training features are calculated in a \texttt{coffea} processor which utilizes \texttt{dask}~\cite{dask} to process multiple chunks of data in parallel.\\ \\
Many machine learning applications in HEP analysis use hyperparameter optimization to identify the best set of hyperparameters to maximize the performance metric of interest. 
For this demonstration, a random search hyperparameter optimization approach is adopted. 
A chosen number of test hyperparameter values are randomly selected from a preset parameter space. 
A BDT model is then created using each set of hyperparameters. 
In this case, since each trial is independent of the previous trial, each model can be trained in parallel. 
This is achieved again using \texttt{dask}. \texttt{mlflow}~\cite{mlflow} is used here as a tracking tool to store different performance metrics for each trial, including the accuracy, precision, and training time. 
In addition, \texttt{mlflow} can also be used to store the trained models so that local memory can be saved. 
This is particularly useful for more complex models.\\ \\
Once the best model is selected from all trials, the best model is saved into \texttt{Amazon s3} storage and from there is connected to the \texttt{NVIDIA Triton} inference server. A configuration file specifying the model name, attributes, and inference preferences is included alongside the models to inform \texttt{Triton} how to load and use the model.

\section{Inference Component}
\begin{figure}[h]
\centering
\includegraphics[width=13cm,clip]{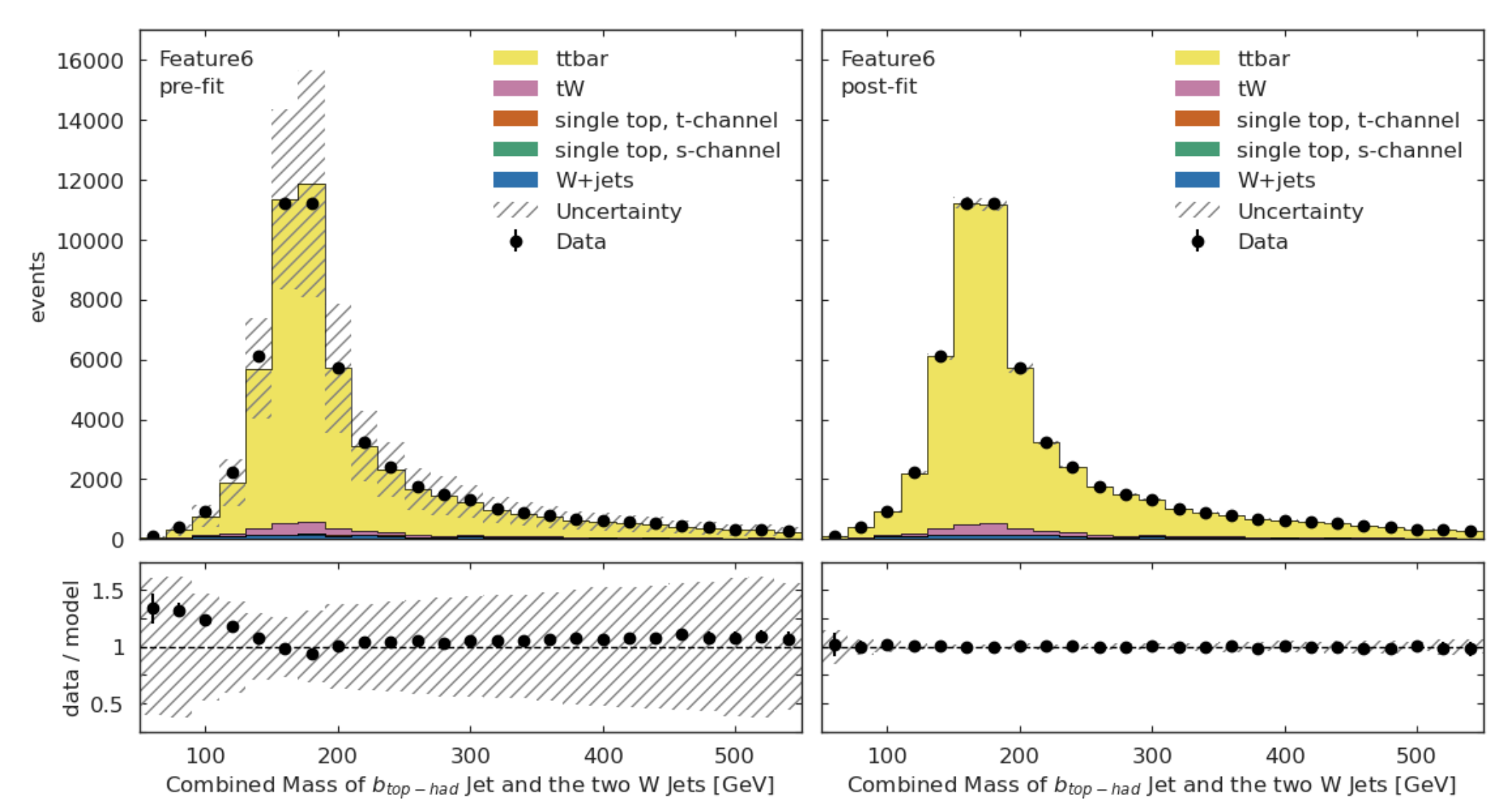}
\caption{The fit results are applied to one of the ML observables (the ML version of the reconstructed top mass). The data-model ratios in the post-fit distribution are all consistent with 1, indicating that the fit works well.}
\label{fit_comparison}
\end{figure}

The inference component of the machine learning task utilizes \texttt{NVIDIA Triton}~\cite{triton} to perform inference as a service. The pre-trained models from the training step are located in \texttt{Amazon s3} storage. In order to perform inference, the user first initializes a \texttt{Triton} gRPC client to communicate with the inference server. Once the input features are calculated (using the same steps as in the training component except using no truth information), the user specifies the model, sends the columns of features to the inference server, then receives the inference results. Since the BDT model is relatively simple, the inference server does not save much time. Time improvements will be more apparent when a more complex model is introduced.\\ \\
After the inference results are obtained, the permutations with the highest score are selected. The combined mass trijet combination in each event with the highest $p_T$ is used as the signal observable, which uses no information from the machine learning component. This observable is an approximation of the top mass. The signal observable is used to fit the model to the data. The fit results can then be applied to the machine learning observables to assess goodness of fit. An example comparing the pre-fit and post-fit distributions of one of the ML observables is seen in Figure \ref{fit_comparison}.
\section{Conclusions and Outlook}
Columnar analysis is an important tool for HEP scientists who want to conduct their analysis in Python. Since ML is an important component of many recent HEP analysis, it is useful to show how ML inference can be integrated into a columnar analysis framework. This initial integration of a machine learning component into the Analysis Grand Challenge framework demonstrates one way of using ML in columnar analysis. Different tools can be used to facilitate the integration of ML at different steps along the process. There are many options available to users, only some of which have been explored so far in the context of the Analysis Grand Challenge.\\ \\
For the training step, distributing different hyperparameter optimization trials using \texttt{dask} is used to speed up the training process. Since we have been working with a simple model thus far, all training is conducted using the local CPU resources. \texttt{mlflow} is used to track and store models outside of the local \texttt{Jupyter} notebook kernel. For inference, \texttt{NVIDIA Triton} is used to perform "inference as a service", which ships columns of input features to perform inference on GPUs and return the column of results.\\ \\
Future studies will involve benchmarking measurements of the inference component in order to identify potential bottlenecks and identify where improvement is necessary. In order to observe the improvements offered by the utilization of the \texttt{NVIDIA Triton} inference server, a more complex model will be developed and used. Once a more complex model is adopted, it will likely be necessary to adapt the approach of training the model to include training on a GPU.

\section*{Acknowledgements}
This work was supported by the U.S. National Science Foundation (NSF) Cooperative Agreement OAC-1836650 (IRIS-HEP).\\ \\
The Analysis Grand Challenge is made possible thanks to the help of a large number of people working on many different projects.
Thank you in particular to the teams behind: coffea-casa, Scikit-HEP, coffea, IRIS-HEP Analysis Systems, ServiceX, IRIS-HEP DOMA, IRIS-HEP SSL, and the CMS Data Preservation and Open Access (DPOA) group.

\bibliography{bib/references}

\end{document}